\documentclass[aps,prd,reprint,groupedaddress,nofootinbib,floatfix]{revtex4-2}

\usepackage{amssymb}
\usepackage{amsthm}
\usepackage{mathtools}
\usepackage{enumitem}
\usepackage{graphicx}
\usepackage{hyperref}
\usepackage{cleveref}
\usepackage[dvipsnames]{xcolor}
\usepackage{tikz}
\usetikzlibrary{positioning}
\usepackage{siunitx}
\usepackage{float}
\usepackage[T1]{fontenc}
\usepackage{booktabs}
\usepackage{natbib}

\hypersetup{
	colorlinks=true,
	linkcolor=blue,  
	urlcolor=blue,
	citecolor=blue
}

\definecolor{mylinkcolor}{RGB}{0,0,0}  

\definecolor{lightred}{rgb}{0.884 0.584 0.619}
\definecolor{darkred}{rgb}{0.363 0 0}
\definecolor{lightblue}{rgb}{0.555 0.693 0.804}
\definecolor{darkblue}{rgb}{0 0 0.363}
\definecolor{lightgreen}{rgb}{0.462 0.835 0.462}
\definecolor{darkgreen}{rgb}{0 0.392 0}

\colorlet{colorvariable}{mylinkcolor}

\begin{document}


\title{Circular geodesic orbits in the equatorial plane of a charged rotating disc of dust}


\author{David Rumler}
\email[]{david.rumler@uni-jena.de}
\affiliation{Theoretisch-Physikalisches Institut, Friedrich-Schiller-Universität Jena, Max-Wien-Platz 1, D-07743 Jena, Germany}


\date{\today}

\begin{abstract}
	Equatorial circular geodesic orbits of neutral test particles in the exterior spacetime of a charged rotating disc of dust are analyzed in dependence of its specific charge and a relativity parameter.
	The charged rotating disc of dust is an axisymmetric, stationary solution of the Einstein-Maxwell equations in terms of a post-Newtonian expansion. In particular, photon, marginally bound and marginally stable orbits are discussed.
	It turns out that general formulae in closed form for angular velocity, specific energy and specific angular momentum of the test particles can be derived, which hold for any (exterior) asymptotically flat, axisymmetric, stationary and reflection symmetric \mbox{(electro-)vacuum} spacetime.
	Furthermore, circular motion in the exterior spacetime of the charged rotating disc of dust is qualitatively very similar to that around a Kerr-Newman black hole for sufficiently large radii, but differs strongly in the respective closer vicinity.
\end{abstract}


\maketitle

\section{Introduction}
\label{sec:introduction}

Investigating the motion of test particles is a very direct and instructive method for analyzing the geometric structure of a given spacetime within the framework of general relativity. It is well known that freely falling neutral test particles travel along geodesics. 

The motion of test particles also plays a role in astrophysics, for instance in the accretion of matter by a central object and in the (indirect) formation of black hole shadows by circular photon orbits.

Various aspects of geodesic motion around Kerr and Kerr-Newman black holes have been studied, see, e.g., the publications by Bardeen et al.\ \cite{Bardeen1972} and Dadhich and Kale \cite{dadhich2008}.\footnote{Also (electrically) charged test particle orbits in the Kerr-Newman spacetime have been explored (see, e.g., \cite{Bicak1989,Balek1989II}). However, those orbits are no longer geodesic. Charged test particles generally move under the influence of the gravitational and the electromagnetic field (if present).}

Meinel and Kleinwächter \cite{Meinel1995dragging} (see also \cite{RFE}) and Ansorg \cite{Ansorg1998} analyzed (timelike) geodesic motion in the gravitational field of an uncharged rotating disc of dust \cite{Neugebauer:1993ct, PhysRevLett.75.3046, RFE} (which may serve as a basic model for disc galaxies).

In the present paper circular orbits of neutral test particles in the equatorial plane of an isolated, axisymmetric, stationary, reflection symmetric, charged, rotating body are investigated first.
This class includes black holes (in particular the Kerr-Newman solution), rotating disc configurations (such as rigidly rotating discs of dust with or without charge) and generally (idealized, uncharged) astrophysical objects.
Remarkably, even without knowing the explicit solution of the rotating body, energy, angular momentum and angular velocity of the test particles can be specified in closed form solely as functions of the metric and their radial derivatives. 

The general examination of equatorial circular orbits of neutral test particles is followed by a specialization on the spacetime of a charged rotating disc of dust. This disc is rigidly rotating, has constant specific charge and is an axisymmetric, stationary solution of the Einstein-Maxwell equations expressed in terms of a post-Newtonian expansion up to tenth order \cite{Palenta_2013,Breithaupt_2015}. A detailed discussion of equatorial circular orbits in the spacetime of the charged rotating disc of dust is provided.

\section{Charged rotating disc of dust}
\label{sec:disc}

Within general relativity an infinitesimally thin equilibrium disc configuration is considered. The disc is made of dust (i.e.\ a perfect fluid with vanishing pressure), it carries a constant specific (electric) charge $\epsilon \in \left[-1,1\right]$ (electric charge density over baryonic mass density) and it is rotating rigidly around the axis of symmetry with a constant angular velocity $\Omega$. It is, furthermore, axisymmetric, stationary and it obeys reflection symmetry \cite{Palenta_2013, Breithaupt_2015, Meinel_2015}. 

Using axisymmetry and stationarity the coupled Einstein-Maxwell equations in electro-vacuum can be reduced to the Ernst equations \cite{PhysRev.168.1415}. Together with reflection symmetry, a well-defined boundary value problem to the Ernst equations for the charged rotating disc of dust can be formulated (see \cite{Palenta_2013,Rumler}). It was solved by means of a post-Newtonian expansion up to eighth order by Palenta and Meinel \cite{Palenta_2013} and up to tenth by Breithaupt et al.\ \cite{Breithaupt_2015}.

The corresponding line element of the disc is expressed globally in terms of Weyl-Lewis-Papapetrou coordinates\footnote{Units with $c=G=4\pi\epsilon_{0}=1$ are used.},
\begin{equation}\label{eq:bvp.4}
	\mathrm{d}s^2 = f^{-1}\left[ h\left( \mathrm{d}\rho^2 + \mathrm{d}\zeta^2 \right) + \rho^2\mathrm{d}\varphi^2 \right] - f\left( \mathrm{d}t + a\, \mathrm{d}\varphi \right)^2 \,,
\end{equation}
and the electromagnetic four-potential is given by
\begin{equation}
	A_a = (0,0,A_{\varphi}, A_t)\,.
\end{equation}
Introducing a relativity parameter $g\coloneqq\sqrt{\gamma}$, where 
\begin{equation}
	\gamma \coloneqq 1 - \sqrt{f_{c}} \,, \quad \text{with} \quad f_{c} \coloneqq f\left(\rho=0, \zeta=0\right) \,,
\end{equation}
was originally used by Bardeen and Wagoner \cite{bardeen}, the disc solution can be specified by:
\begin{align}
	&f = 1 + \sum_{k=1}^{10}f_{2k}g^{2k} \,, \quad h = 1 + \sum_{k=2}^{10}h_{2k}g^{2k} \,, \label{eq:pne1}\\
	&a^* = \sum_{k=1}^{10}a^*_{2k+1}g^{2k+1} \,, \label{eq:pne2}\\
	&A_{\varphi}^* = \sum_{k=1}^{10}A_{\varphi\,2k+1}^{*}g^{2k+1} \,, \quad A_{t} = \sum_{k=1}^{10}A_{t\,2k}g^{2k} \label{eq:pne3}\,.
\end{align}
The coefficient functions $f_{2k}$, $h_{2k}$, $a^*_{2k+1}$, ... only depend on the elliptic coordinates $\eta\in[-1,1]$ and $\nu\in[0,\infty]$, which are defined via
\begin{equation}
	\rho = \rho_0\sqrt{\left(1-\eta^2\right)(1+\nu^2)} \,, \quad \zeta=\rho_0\eta\nu \,.
\end{equation}
Here, * denotes the normalization by suitable powers of the disc's coordinate radius $\rho_{0}$ in order to obtain dimensionless quantities, such as $a^*=\frac{a}{\rho_{0}}$ and $A_{\varphi}^{*}=\frac{A_{\varphi}}{\rho_{0}}$. 

Note that the relativity parameter can equivalently be expressed in terms of the redshift $z_{c}$ of a photon traveling from the center of the disc to spatial infinity:
\begin{equation}
	g^2 = \frac{z_{c}}{1+z_{c}} \,.
\end{equation}
Therefore, the relativity parameter connects the regime of Newtonian physics, $g\ll 1$, with the ultra-relativistic regime, $g \to 1$, (where black hole formation is assumed) and  
the parameter space of the disc solution is spanned by $g\in[0,1]$, $\epsilon \in [0,1]$ and $\rho_{0}$. Without loss of generality only positive charges are considered and the coordinate radius $\rho_{0}$ serves as a scaling parameter. All physical quantities studied in this paper are functions of $g$, $\epsilon$ and $\rho_{0}$ only.

The specific charge $\epsilon$, furthermore, directly influences the disc's rotation speed.
As is evident from the Newtonian limit (where each dust particle within the disc is in an equilibrium of gravitational, electric and centrifugal force), disc configurations with $\epsilon=0$ have a maximum angular velocity and those with $\epsilon=1$ are static.\footnote{A detailed investigation of the dependence of the gravitational mass $M$, the angular momentum $J$, the electric charge $Q$ and the magnetic dipole moment $D$ (each with an appropriate normalization), as well as higher (normalized) multipole moments, on the parameters $g$ and $\epsilon$ can be found in \cite{Rumler_MultipoleMoments}.}

\section{Circular geodesic orbits in a general spacetime}\label{sec:orbits.uncharged}

We first consider the exterior spacetime of an isolated, axisymmetric, stationary, reflection symmetric, charged, rotating body. These qualifications, besides a non-zero charge, are expected to be fulfilled for most astrophysical bodies.\footnote{Axisymmetry, stationarity and reflection symmetry should be fulfilled to a good approximation for most astrophysical objects.} (Note that the special cases of zero charge and vanishing rotation are included.)

A freely falling uncharged test particle moves through spacetime along a (timelike) curve determined by the geodesic equation,
\begin{equation}\label{eq:cm.geodesiceq}
	\frac{\mathrm{d}^2x^{i}}{\mathrm{d}\tau^2} + \Gamma^{i}_{mn}\frac{\mathrm{d}x^{m}}{\mathrm{d}\tau}\frac{\mathrm{d}x^{n}}{\mathrm{d}\tau} = 0 \,,
\end{equation} 
where $\tau$ is the proper time of the test particle and $\Gamma^{i}_{mn}$ are the Christoffel symbols.

The body is arranged so that the axis of rotation coincides with the $\zeta$-axis and its equatorial plane is defined by $\zeta=0$. We restrict the motion of the test particles to this plane, i.e.\ $\zeta=0$, $\frac{\mathrm{d}\zeta}{\mathrm{d}\tau}=0$  and $\frac{\mathrm{d}^2\zeta}{\mathrm{d}\tau^2}=0$. (In order to keep the notation simple, the dependency on the coordinates is suppressed and, e.g., $g_{ab}$  is written instead of $g_{ab}(\rho,\zeta=0)$ in the following.) Let the geodesics, furthermore, be circular orbits.

Due to stationarity and axisymmetry the metric $g_{ab}$ of the exterior spacetime of the body can be expressed in terms of Weyl-Lewis-Papapetrou coordinates, see \cref{eq:bvp.4}.

The circular motion of a test particle with constant angular velocity $\Omega_{\text{tp}}=\frac{\mathrm{d}\varphi}{\mathrm{d}t}$ (as seen from infinity) is also stationary and the four-velocity is therefore, given by
\begin{align}
	\left(u_{\text{tp}}\right)^{i}&=\frac{\mathrm{d}x^{i}}{\mathrm{d}\tau}=\left(0,0,\Omega_{\text{tp}}e^{-U'},e^{-U'}\right) \,, \label{eq:cm.4velocity} \\
 	\left(u'_{\text{tp}}\right)^{i}&=\left(0,0,0,e^{-U'}\right) \,, \label{eq:cm.4velocity2}
\end{align}
where $f=e^{2U}$ and the primes denote the frame of reference co-rotating to the test particle, with $\varphi' = \varphi - \Omega_{\text{tp}}t$.\footnote{Note that in the literature listed on the uncharged and charged rotating disc of dust, the primes refer to the co-rotating frame of the rigidly rotating disc. Notation-wise this, however, does not cause a problem, as no reference to the co-rotating frame of the disc is made here.}

Using \cref{eq:cm.4velocity2}, the only non-vanishing component of the geodesic equation (\ref{eq:cm.geodesiceq}) is obtained for $i=\rho$:
\begin{equation}\label{eq:cm.eom}
	g'_{tt,\rho} = 0 \,.
\end{equation}
The $\zeta$-component vanishes due to the restriction of the motion to the equatorial plane of the body and reflection symmetry of the metric. In other words, setting $\zeta=0$, $\frac{\mathrm{d}\zeta}{\mathrm{d}\tau}=0$  and $\frac{\mathrm{d}^2\zeta}{\mathrm{d}\tau^2}=0$ is actually solving the $\zeta$-component of the geodesic equation.

Note that the transformation law for a transition to the co-rotating frame of the test particle reads in case of $g_{tt}$:
\begin{equation}\label{eq:cm.gtttransf}
	g'_{tt} = g_{tt} + 2\Omega_{\text{tp}}g_{\varphi t} + \Omega_{\text{tp}}^{2}g_{\varphi\varphi} \,.
\end{equation}
Eq.\ (\ref{eq:cm.eom}) is therefore equivalent to 
\begin{equation}
	g_{\varphi\varphi,\rho}\Omega_{\text{tp}}^2 + 2g_{\varphi t,\rho}\Omega_{\text{tp}} + g_{tt,\rho} = 0 \,.
\end{equation}
This immediately gives the solution for the angular velocity of the test particle:
\begin{equation}\label{eq:cm.Omega}
	\Omega_{1/2} = \frac{-g_{\varphi t,\rho} \pm \sqrt{g_{\varphi t,\rho}^{2} - g_{\varphi\varphi,\rho}g_{tt,\rho}}}{g_{\varphi\varphi,\rho}} \,,
\end{equation}
where (in general) one of the solutions corresponds to a prograde and the other one to a retrograde orbit.

The two symmetries, axisymmetry and stationarity, give rise to two conserved quantities associated with the motion of the test particle. Axisymmetry represented by the spacelike Killing vector $\boldsymbol{\eta}=\frac{\partial}{\partial \varphi}$  implies a conserved specific angular momentum \mbox{$L=\eta_{i}\left(u_{\text{tp}}\right)^{i}$} and stationarity with a timelike Killing vector $\boldsymbol{\xi}=\frac{\partial}{\partial t}$ ensures the conservation of the specific energy $E=-\xi_{i}\left(u_{\text{tp}}\right)^{i}$.
With the four-velocity from \cref{eq:cm.4velocity} one obtains the specific angular momentum
\begin{equation}\label{eq:cm.L}
	L_{1/2} = \frac{g_{\varphi\varphi}\Omega_{1/2}+g_{\varphi t}}{\sqrt{- g_{\varphi\varphi}\Omega_{1/2}^{2} -2g_{\varphi t}\Omega_{1/2} - g_{tt}}}
\end{equation}
and the specific energy
\begin{equation}\label{eq:cm.E}
	E_{1/2} = -\frac{g_{\varphi t}\Omega_{1/2}+g_{tt}}{\sqrt{- g_{\varphi\varphi}\Omega_{1/2}^{2} -2g_{\varphi t}\Omega_{1/2} - g_{tt}}}
\end{equation}
of the test particle. 
Prograde orbits are defined by the positive specific angular momentum solution and retrograde orbits by the negative one.

It should be stressed that \cref{eq:cm.Omega,eq:cm.L,eq:cm.E} hold for any (exterior) asymptotically flat, axisymmetric, stationary and reflection symmetric \mbox{(electro-)vacuum} spacetime and the solutions could be derived in a closed form without knowledge of the concrete spacetime metric.

Circular orbits only exist for
\begin{equation}
	g_{\varphi\varphi}\Omega_{1/2}^2 + 2g_{\varphi t}\Omega_{1/2} + g_{tt} \leq 0 \,,
\end{equation}
according to \cref{eq:cm.L,eq:cm.E}. In the limiting case of equality, specific angular momentum and specific energy diverge. This characterizes a photon orbit for prograde and retrograde motion respectively. Prograde or retrograde circular motion of uncharged test particles is therefore only possible for radii larger than that of the associated photon orbit.\footnote{For charged test particles this innermost boundary of circular orbits generally no longer coincides with the photon orbit (as a result of the electromagnetic interaction between the rotating body and the test particle), see e.g.\ \cite{Balek1989II}.}

There is an equivalent way of deriving \cref{eq:cm.Omega,eq:cm.L,eq:cm.E} via the Lagrange formalism. Generally, the Lagrangian describing the (timelike) motion of a test particle in a spacetime furnished with the metric $g_{ab}$ is
\begin{equation}
	\mathcal{L} = \frac{1}{2}g_{ab}\dot{x}^{a}\dot{x}^{b} \,,
\end{equation}
where $\dot{x}^{a}=\frac{\mathrm{d}x^{a}}{\mathrm{d}\tau}$. The resulting Euler-Lagrange equation is equivalent to the geodesic equation, i.e.\ the extremal path between two points is also the straightest possible path.

As the Lagrangian corresponding to the above defined rotating body does not depend on $\varphi$ and $t$ ($g_{ab}$ is axisymmetric and stationary), there are two conserved quantities:
\begin{align}
	L &\coloneqq \frac{\partial \mathcal{L}}{\partial \dot{\varphi}} = g_{\varphi\varphi}\dot{\varphi} + g_{\varphi t}\dot{t} \,, \label{eq:cm.L2}\\
	E &\coloneqq -\frac{\partial \mathcal{L}}{\partial \dot{t}} = -g_{\varphi t}\dot{\varphi} - g_{tt}\dot{t} \,. \label{eq:cm.E2}
\end{align}
$L$ and $E$ coincide with the specific angular momentum and the specific energy, defined previously via the Killing vectors.

From \cref{eq:cm.L2,eq:cm.E2}, the angular velocity of the test particle $\Omega_{\text{tp}}=\frac{\mathrm{d}\varphi}{\mathrm{d}t}$ (as seen from infinity) can be deduced:
\begin{equation}\label{eq:cm.Omega2}
	\Omega_{\text{tp}} = -\frac{g_{tt}L + g_{\varphi t}E}{g_{\varphi t}L + g_{\varphi\varphi}E} \,.
\end{equation}

As above, we restrict the motion of the test particle to the equatorial plane of the body.
Inserting $\dot{\varphi}$ and $\dot{t}$, using \cref{eq:cm.L2,eq:cm.E2}, into the normalization condition of the four-velocity for timelike curves, $g_{ab}\dot{x}^{a}\dot{x}^{b} = -1$, leads to
\begin{equation}\label{eq:cm.effectivepot}
	\frac{1}{2}g_{\rho\rho}\left(\frac{\mathrm{d}\rho}{\mathrm{d}\tau}\right)^{\!2} + \mathcal{U} = 0 \,,
\end{equation}
where
\begin{equation}\label{eq:cm.effectivepot2}
	\mathcal{U} \coloneqq -\frac{1}{2\rho^{2}}\left(g_{tt}L^{2} +2g_{\varphi t}LE + g_{\varphi\varphi}E^2\right) + \frac{1}{2} 
\end{equation}
is an effective potential. The motion of the test particle in the equatorial plane can therefore also be described by a motion in an effective potential.

A useful relation that was applied to $\mathcal{U}$ and that is also employed in subsequent calculations is
\begin{equation}
	g_{\varphi t}^{2} - g_{\varphi\varphi}g_{tt} = \rho^2 \,.
\end{equation}

Differentiation of \cref{eq:cm.effectivepot} with respect to the particle proper time $\tau$ leads to the equation of motion:
\begin{equation}\label{eq:cm.eom2}
	g_{\rho\rho}\frac{\mathrm{d}^2\rho}{\mathrm{d}\tau^2} + \frac{1}{2}g_{\rho\rho,\rho}\left(\frac{\mathrm{d}\rho}{\mathrm{d}\tau}\right)^{\!2} + \mathcal{U}_{,\rho} = 0 \,.
\end{equation}
One can verify that \cref{eq:cm.eom2} corresponds to the $\rho$-component of the geodesic equation.\footnote{For not necessarily circular motion, the $\rho$-component of the geodesic equation is given by: \newline $g_{\rho\rho}\frac{\mathrm{d}^2\rho}{\mathrm{d}\tau^2} + \frac{1}{2}g_{\rho\rho,\rho}\left(\frac{\mathrm{d}\rho}{\mathrm{d}\tau}\right)^{\!2} - \frac{1}{2}\left(g_{\varphi\varphi,\rho}\Omega_{\text{tp}}^2 + 2g_{\varphi t,\rho}\Omega_{\text{tp}} + g_{tt,\rho}\right)\left(\frac{\mathrm{d}t}{\mathrm{d}\tau}\right)^{\!2} \\ = 0$.}

Circular orbits must fulfill the requirements $\frac{\mathrm{d}\rho}{\mathrm{d}\tau}=0$ and $\frac{\mathrm{d}^2\rho}{\mathrm{d}\tau^2}=0$. According to \cref{eq:cm.effectivepot,eq:cm.eom2}, the conditions for circular orbits (in the equatorial plane) are therefore $\mathcal{U}=0$ and $\mathcal{U_{,\rho}}=0$.
These conditions imply:
\begin{align}
	&g_{tt}L^{2} + 2g_{\varphi t}LE + g_{\varphi\varphi}E^2 - \rho^{2} = 0 \,, \label{eq:cm.effcond1} \\
	&g_{tt,\rho}L^{2} + 2g_{\varphi t,\rho}LE + g_{\varphi\varphi,\rho}E^2 - 2\rho = 0 \,. \label{eq:cm.effcond2}
\end{align} 
Notice that the above equations agree with those derived in \cite{Ansorg1998}.

Solving \cref{eq:cm.effcond1,eq:cm.effcond2} reveals the specific angular momentum and the specific energy of the test particle:
\begin{align}
	\left\vert L_{1/2} \right\vert &= \left\vert l_{1/2} \right\vert \frac{\rho}{\sqrt{g_{tt}l_{1/2}^2 + 2g_{\varphi t}l_{1/2} + g_{\varphi\varphi}}} \,, \label{eq:cm.Lm2}\\
	\left\vert E_{1/2} \right\vert &= \frac{\rho}{\sqrt{g_{tt}l_{1/2}^2 + 2g_{\varphi t}l_{1/2} + g_{\varphi\varphi}}} \,, \label{eq:cm.Em2}
\end{align}
where
\begin{equation}\label{eq:cm.l}
	l_{1/2} \coloneqq \frac{L_{1/2}}{E_{1/2}} = - \frac{2g_{\varphi t} -\rho g_{\varphi t,\rho} \pm \rho\sqrt{g_{\varphi t,\rho}^{2} - g_{\varphi\varphi,\rho}g_{tt,\rho}}}{2g_{tt} - \rho g_{tt,\rho}} \,.
\end{equation}
The sign of $E_{1/2}$ follows from \cref{eq:cm.E2} (where \mbox{$\dot{t}>0$}) and the sign of $L_{1/2}$ is then automatically given by \mbox{$L_{1/2}=l_{1/2}E_{1/2}$}.
Clearly, the angular velocity (see \cref{eq:cm.Omega2}) can also be expressed in terms of $l_{1/2}$:
\begin{equation}\label{eq:cm.Omegam2}
	\Omega_{1/2} = -\frac{g_{tt}l_{1/2} + g_{\varphi t}}{g_{\varphi t}l_{1/2} + g_{\varphi\varphi}} \,.
\end{equation}

It can be shown explicitly that the two sets of equations for specific angular momentum, specific energy and angular velocity, i.e.\ \cref{eq:cm.Omega,eq:cm.L,eq:cm.E} and \cref{eq:cm.Omegam2,eq:cm.Lm2,eq:cm.Em2,eq:cm.l}, are indeed equivalent. In particular, the following relation holds:
\begin{align}
	&g_{tt}l_{1/2}^2 + 2g_{\varphi t}l_{1/2} + g_{\varphi\varphi} \notag \\
	&= \frac{-\rho^{2}}{\left(-g_{\varphi t}\Omega_{1/2} - g_{tt}\right)^2}\left(g_{\varphi\varphi}\Omega_{1/2}^{2} + 2g_{\varphi t}\Omega_{1/2} + g_{tt}\right) \,.
\end{align}
Eq.\ (\ref{eq:cm.Omegam2}) can furthermore be easily converted to
\begin{equation}
	l_{1/2} = - \frac{g_{\varphi\varphi}\Omega_{1/2} + g_{\varphi t}}{g_{\varphi t}\Omega_{1/2} + g_{tt}} \,.
\end{equation}

The advantage of the second approach (using the Lagrange formalism in conjunction with the effective potential) is that the stability of the circular orbits can be analyzed directly on the basis of the effective potential. According to \cref{eq:cm.effectivepot}, non-circular motion is only possible for $\mathcal{U}<0$. For given $L$ and $E$, circular orbits are therefore stable (with respect to radial and angular perturbations) if and only if the extremum of $\mathcal{U}$ is a minimum, i.e.\ $\mathcal{U}_{,\rho\rho}>0$. Using $\mathcal{U}=0$ and $\mathcal{U}_{,\rho}=0$, $\mathcal{U}_{,\rho\rho}>0$ is equivalent to
\begin{equation}
	g_{tt,\rho\rho}L^{2} + 2g_{\varphi t,\rho\rho}LE + g_{\varphi\varphi,\rho\rho}E^2 - 2 < 0 \,.
\end{equation}
For $\mathcal{U}_{,\rho\rho}<0$ circular orbits are thus unstable.\footnote{A discussion of the stability of circular orbits in stationary, axisymmetric spacetimes can also be found in \cite{Bardeen_StableOrbits} and \cite{Beheshti}.} 
Moreover, if $E^2<1$ circular orbits are bound, i.e.\ the test particle cannot reach infinity.\footnote{For $E^{2} > 1$ there can also be additional bound orbits, see, e.g., \cite{Ansorg1998}.}

Inserting the metric functions of the uncharged rotating disc of dust \cite{Neugebauer:1993ct, PhysRevLett.75.3046, RFE} at the rim into the above formulae (of the first or second approach) reveals the same expressions for $\Omega_{1/2}$, $L_{1/2}$ and $E_{1/2}$ (with $1/2=+/-$) as in \cite{Meinel1995dragging}.
Also the corresponding results for equatorial circular geodesic orbits in the (exterior) Kerr \cite{Bardeen1972} and Kerr-Newman spacetime \cite{dadhich2008} can be obtained from the above formulae (using either approach).\footnote{Note that $\rho\ge 0$.}

\section{Circular geodesic orbits in the spacetime of a charged rotating disc of dust}\label{sec:orbits.uncharged2}

\begin{figure}[htb]
	\centering
	\begin{minipage}{0.45\textwidth}
		\centering
		\includegraphics[scale=1]{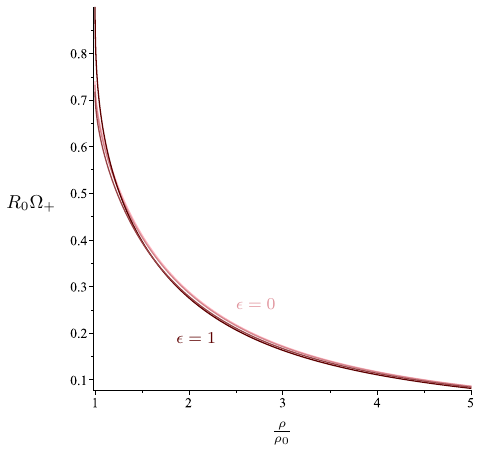}
	\end{minipage}\hfill
	\begin{minipage}{0.45\textwidth}
		\centering
		\includegraphics[scale=1]{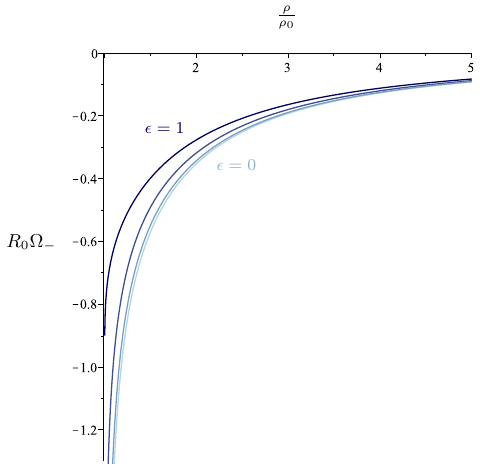}
	\end{minipage}
	\caption{Normalized angular velocities $R_{0}\Omega_{+}$ (top) and $R_{0}\Omega_{-}$ (bottom) of test particles moving along (prograde/retrograde) circular geodesic orbits, shown for $g=0.7$ and \mbox{$\epsilon\in\{0,1/3,2/3,1\}$}.}
	\label{fig:cm.Omegaeps}
\end{figure}

\begin{figure}[tb]
	\centering
	\begin{minipage}{0.45\textwidth}
		\centering
		\includegraphics[scale=1]{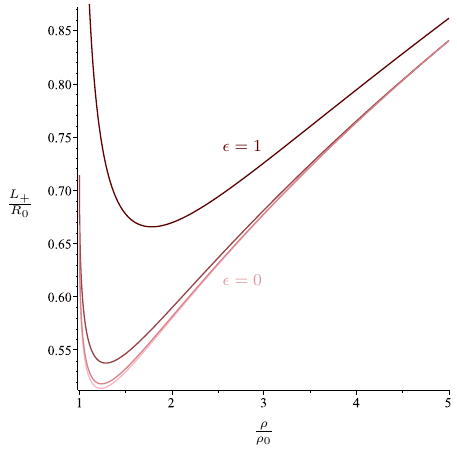}
	\end{minipage}\hfill
	\begin{minipage}{0.45\textwidth}
		\centering
		\includegraphics[scale=1]{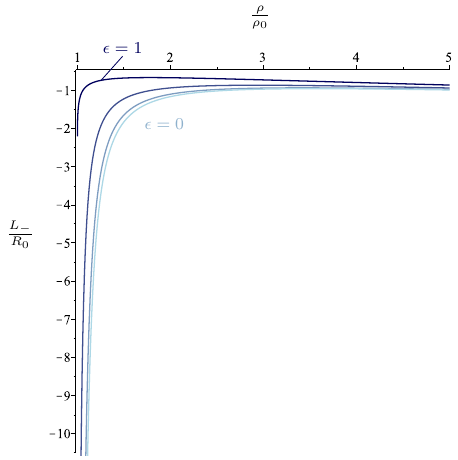}
	\end{minipage}
	\caption{Normalized specific angular momenta $\frac{L_{+}}{R_{0}}$ (top) and $\frac{L_{-}}{R_{0}}$ (bottom) of test particles moving along (prograde/retrograde) circular geodesic orbits, shown for $g=0.7$ and $\epsilon\in\{0,1/3,2/3,1\}$.}
	\label{fig:cm.Leps}
\end{figure}

\begin{figure}[htb]
	\centering
	\begin{minipage}{0.45\textwidth}
		\centering
		\includegraphics[scale=1]{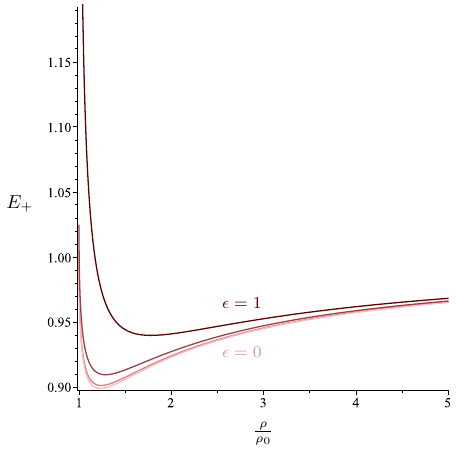}
	\end{minipage}\hfill
	\begin{minipage}{0.45\textwidth}
		\centering
		\includegraphics[scale=1]{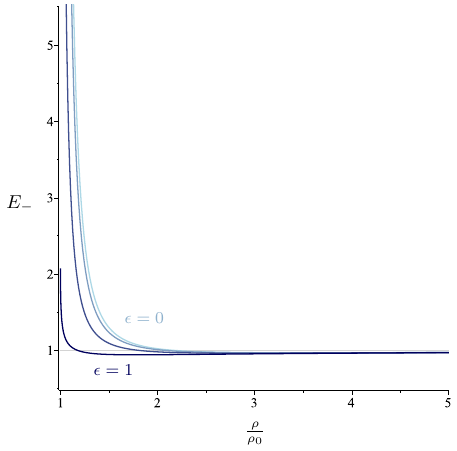}
	\end{minipage}
	\caption{Specific energies $E_{+}$ (top) and $E_{-}$ (bottom) of test particles moving along (prograde/retrograde) circular geodesic orbits, shown for $g=0.7$ and $\epsilon\in\{0,1/3,2/3,1\}$.}
	\label{fig:cm.Eeps}
\end{figure}

\begin{figure}[htb]
	\centering
	\begin{minipage}{0.45\textwidth}
		\centering
		\includegraphics[scale=1]{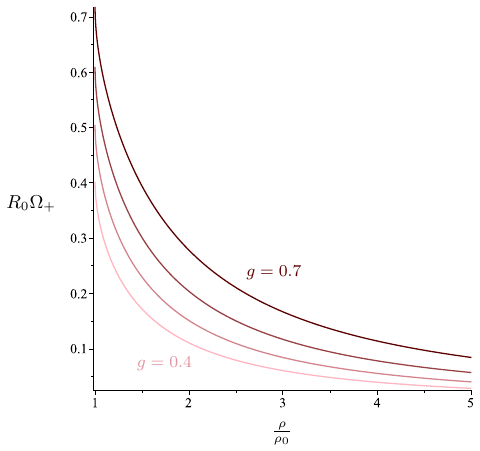}
	\end{minipage}\hfill
	\begin{minipage}{0.45\textwidth}
		\centering
		\includegraphics[scale=1]{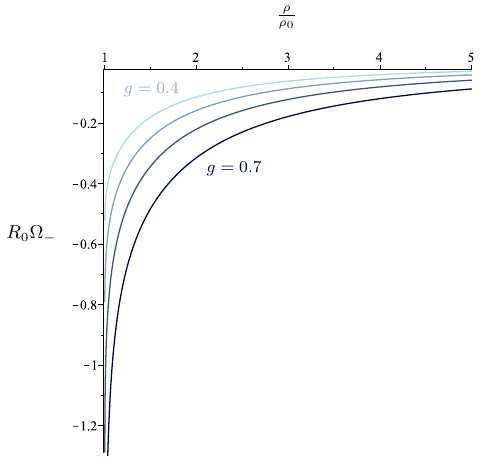}
	\end{minipage}
	\caption{Normalized angular velocities $R_{0}\Omega_{+}$ (top) and $R_{0}\Omega_{-}$ (bottom) of test particles moving along (prograde/retrograde) circular geodesic orbits, shown for $\epsilon=0.7$ and \mbox{$g\in\{0.4,0.5,0.6,0.7\}$}.}
	\label{fig:cm.Omegag}
\end{figure}

\begin{figure}[htb]
	\centering
	\begin{minipage}{0.45\textwidth}
		\centering
		\includegraphics[scale=1]{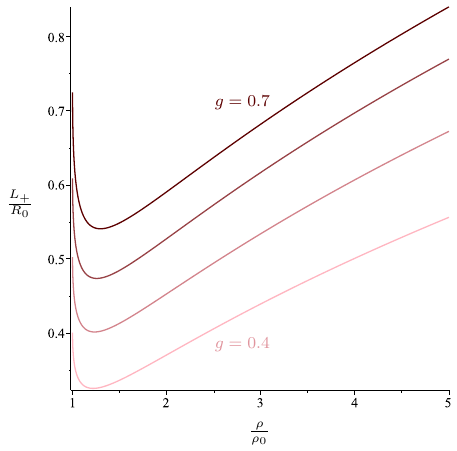}
	\end{minipage}\hfill
	\begin{minipage}{0.45\textwidth}
		\centering
		\includegraphics[scale=1]{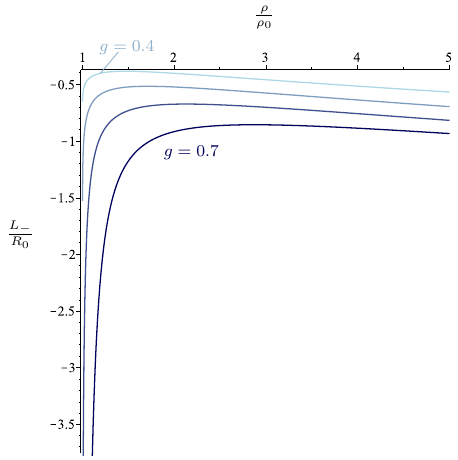}
	\end{minipage}
	\caption{Normalized specific angular momenta $\frac{L_{+}}{R_{0}}$ (top) and $\frac{L_{-}}{R_{0}}$ (bottom) of test particles moving along (prograde/retrograde) circular geodesic orbits, shown for $\epsilon=0.7$ and $g\in\{0.4,0.5,0.6,0.7\}$.}
	\label{fig:cm.Lg}
\end{figure}

\begin{figure}[htb]
	\centering
	\begin{minipage}{0.45\textwidth}
		\centering
		\includegraphics[scale=1]{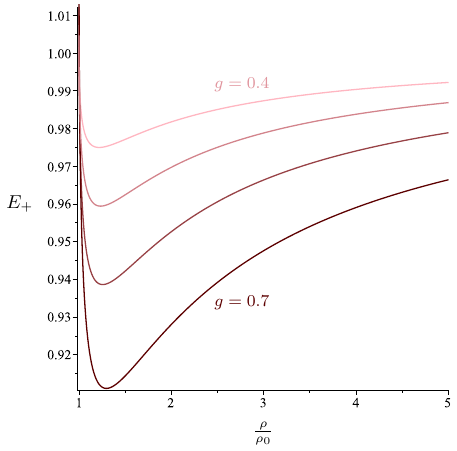}
	\end{minipage}\hfill
	\begin{minipage}{0.45\textwidth}
		\centering
		\includegraphics[scale=1]{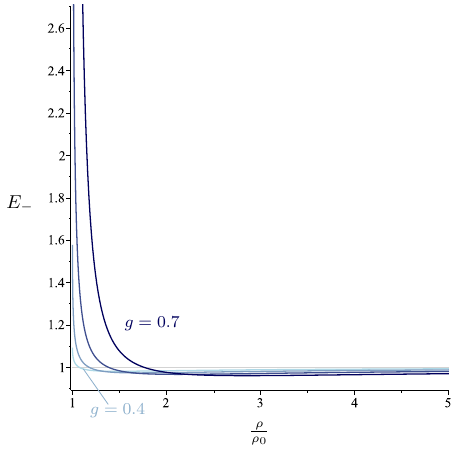}
	\end{minipage}
	\caption{Specific energies $E_{+}$ (top) and $E_{-}$ (bottom) of test particles moving along (prograde/retrograde) circular geodesic orbits, shown for $\epsilon=0.7$ and \mbox{$g\in\{0.4,0.5,0.6,0.7\}$}.}
	\label{fig:cm.Eg}
\end{figure}

Now circular geodesic orbits in the equatorial plane ($\eta=0$) of the charged rotating disc of dust, described in \cref{sec:disc}, are to be discussed. By evaluating \cref{eq:cm.Omega,eq:cm.L,eq:cm.E} or  \cref{eq:cm.Omegam2,eq:cm.Lm2,eq:cm.Em2,eq:cm.l} with the metric of the disc, using \cref{eq:pne1,eq:pne2}, the angular velocity $\Omega_{1/2}=\Omega_{+/-}$, the specific angular momentum $L_{1/2}=L_{+/-}$ and the specific energy \mbox{$E_{1/2}=E_{+/-}$} can be expressed in terms of a post-Newtonian expansion, see \cref{secB}. Here ``$+$'' denotes direct and ``$-$'' retrograde orbits. The angular velocity of the disc, $\Omega$, is non-negative.

It should be noted that the angular velocity $\Omega_{+/-}$ and the specific angular momentum $L_{+/-}$ are odd functions and the specific energy $E_{+/-}$ is an even function in $g$. 
This ensures that $\Omega_{+/-}$ and $L_{+/-}$ change their sign and $E_{+/-}$ remains unchanged under a change of sense of rotation of the disc (for $g\to -g$).\footnote{For further (general) details, see \cite{Palenta_2013} and \cite{Rumler}.}

In the Newtonian limit (i.e.\ up to and including $\mathcal{O}\left(g^2\right)$) prograde and retrograde orbits coincide, as the angular momentum of disc, $J$, does not contribute to the Newtonian gravitational field.\footnote{Indeed, the angular momentum of the disc starts at order $g^3$, see \cite{Rumler_MultipoleMoments}.} In particular, it holds $\Omega_{+}=-\Omega_{-}$, $L_{+}=-L_{-}$ and $E_{+}=E_{-}$. 

Since the angular momentum of the disc vanishes in the limiting case of $\epsilon =1$ (see \cite{Rumler_MultipoleMoments}), prograde and retrograde orbits also coincide for $\epsilon =1$.

It was further confirmed that the simultaneous limit $\epsilon \to 0$, $\rho\to\rho_{0}$ leads to the same results for $\Omega_{\pm}$, $L_{\pm}$ and $E_{\pm}$, in terms of a series expansion in $g$, as in \cite{Meinel1995dragging}. 
Especially noteworthy is the fact that $\Omega_{+}=\Omega$ in this limiting case. 
The reason is that the \mbox{$\rho$-component} of the geodesic equation of a test particle, \cref{eq:cm.eom}, also represents a boundary condition of the uncharged rotating disc of dust (see \cite{RFE}).

Figs.\ \ref{fig:cm.Omegaeps} to \ref{fig:cm.Eg} show, for a fixed proper disc radius $R_{0} \coloneqq \int_{0}^{\rho_{0}}\!\sqrt{g_{\rho\rho}}\,\mathrm{d}\rho$ (see also \cite{Rumler}), the radial dependence of the angular velocity $\Omega_{\pm}$, the specific angular momentum $L_{\pm}$ and the specific energy $E_{\pm}$ for different values of the specific charge $\epsilon$ and the relativity parameter $g$. 
Apart from the region directly adjacent to the rim of the disc, the qualitative behavior of the curves is very similar to the ones corresponding to the (exterior) spacetime of a Kerr-Newman black hole.

As can be seen in \cref{fig:cm.Omegaeps,fig:cm.Leps,fig:cm.Eeps}, reducing the specific charge $\epsilon$ from $1$ to $0$ results (for fixed $R_{0}$) in a decrease in the absolute values of $L_{+}$ and $E_{+}$ and an increase in the absolute values of $\Omega_{-}$, $L_{-}$ and $E_{-}$. $\Omega_{+}$ shows a mixed behavior (for sufficiently large radii, however, it increases). Furthermore, raising the angular momentum of a Kerr-Newman black hole leads to the same qualitative change of the absolute values as for the charged rotating disc of dust by decreasing $\epsilon$ (with the exception of $\Omega_{+}$). This is plausible, since the angular momentum of the disc comes with the global prefactor $\sqrt{1-\epsilon^2}$, see \cite{Rumler_MultipoleMoments}.\footnote{Increasing the charge parameter of the Kerr-Newman black hole qualitatively coincides with an increase of the specific charge of the disc in case of $\Omega_{-}$, $L_{-}$ and $E_{-}$ (as well as $\Omega_{+}$ for sufficiently large radii).}

On the other hand, the absolute values of $R_{0}\Omega_{\pm}$ and $\frac{L_{\pm}}{R_{0}}$ increase and $E_{+}$ decreases with the relativity parameter $g$, see \cref{fig:cm.Omegag,fig:cm.Lg,fig:cm.Eg}. $E_{-}$ increases close to the rim and decreases for sufficiently large radii.

\begin{figure}[tb]
	\centering
	\begin{minipage}{0.45\textwidth}
		\centering
		\includegraphics[scale=1]{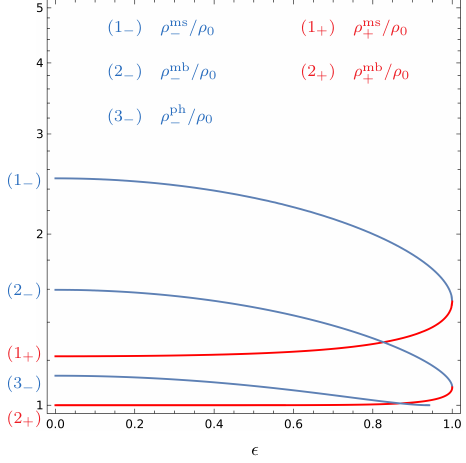}
	\end{minipage}\hfill
	\vspace{0.3cm}
	\begin{minipage}{0.45\textwidth}
		\centering
		\includegraphics[scale=1]{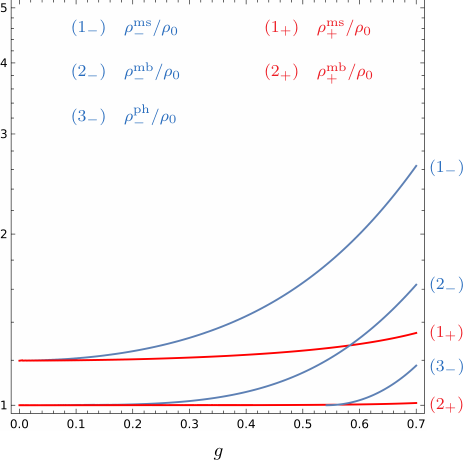}
	\end{minipage}
	\caption{Radii of photon, marginally bound and marginally stable orbits, $\frac{\rho_{-}^{\text{ph}}}{\rho_{0}}$, $\frac{\rho_{\pm}^{\text{mb}}}{\rho_{0}}$ and $\frac{\rho_{\pm}^{\text{ms}}}{\rho_{0}}$, as functions of $\epsilon$ (top), for $g=0.6$, and $g$ (bottom), for $\epsilon=0.8$.}
	\label{fig:cm.phbdst}
\end{figure}

Interesting in the context of (equatorial) circular motion of neutral test particles are also the radii of photon orbits ($g'_{tt}=0$), marginally bound orbits ($E^{2}=1$) and marginally stable orbits ($\mathcal{U}_{,\rho\rho}=0$). They are shown as functions of $\epsilon$ and $g$ in \cref{fig:cm.phbdst}.\footnote{Note that circular orbits with radii smaller than that of the photon orbit / marginally bound orbit / marginally stable orbit are forbidden / unbound / unstable and those with larger radii are permitted / bound / stable.} The corresponding radii of the retrograde orbits decrease and those of the prograde orbits increase with $\epsilon\in \left[0,1\right]$. In the case of $\epsilon=1$ they coincide. For growing $g$ both the radii of retrograde and prograde orbits increase and they merge in the Newtonian limit ($g\ll1$). For sufficiently large values of $\epsilon$ and $g$ there are both retrograde and prograde photon, marginally bound and marginally stable orbits. However, in contrast to the (exterior) spacetime of a Kerr-Newman black hole, for sufficiently small $g$ there are no prograde photon orbits (as shown in \cref{fig:cm.phbdst}) and for even smaller $g$ there are also no retrograde photon orbits (each depending on $\epsilon$).\footnote{In the Newtonian theory of gravity there are in general no photon orbits.} In other words, there are disc configurations for which both prograde and retrograde circular motion is possible arbitrarily close to the rim of the disc, i.e.\ no orbits are forbidden. For $g\to0$ all orbits are bound. Moreover, in comparison to the uncharged rotating disc of dust, for $\epsilon>0$ not all prograde orbits are bound (and for sufficiently large $\epsilon$ and $g$ there are prograde photon orbits). Note that for $g=0.7$ there is no ergosphere yet.

\begin{figure}[tb]
	\centering
	\begin{minipage}{0.45\textwidth}
		\centering
		\includegraphics[scale=1]{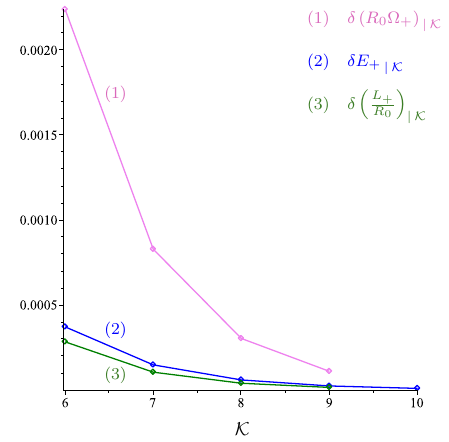}
	\end{minipage}\hfill
	\begin{minipage}{0.45\textwidth}
		\centering
		\includegraphics[scale=1]{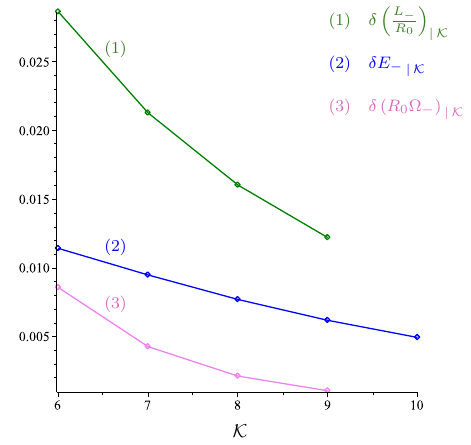}
	\end{minipage}
	\caption{Convergence estimate of the functions $R_{0}\Omega_{+}$, $\frac{L_{+}}{R_{0}}$, $E_{+}$ (top) and $R_{0}\Omega_{-}$, $\frac{L_{-}}{R_{0}}$, $E_{-}$ (bottom), displayed for $g=0.7$, $\epsilon=0$ and $\frac{\rho}{\rho_{0}}=2$.}
	\label{fig:cm.Conv}
\end{figure}

Finally, the convergence behavior of the normalized angular velocity $R_{0}\Omega_{\pm}$, the normalized specific angular momentum $\frac{L_{\pm}}{R_{0}}$ and the specific energy $E_{\pm}$ is illustrated in \cref{fig:cm.Conv}. 
As an example, to estimate the convergence of the specific energy, the resulting change by adding the $\mathcal{K}$th order to the post-Newtonian expansion at order $\mathcal{K}-1$ relative to the expansion solution at $\mathcal{K}$th order is calculated: 
\begin{align}\label{eq:conv}
	\delta {E_{\pm}}_{\,\vert\,\mathcal{K}} \coloneqq \frac{\left\vert {E_{\pm}}_{\,\vert\,\mathcal{K}} - {E_{\pm}}_{\,\vert\,\mathcal{K}-1}\right\vert}{\left\vert {E_{\pm}}_{\,\vert\,\mathcal{K}}\right\vert} \,, \\
	\text{where} \quad {E_{\pm}}_{\,\vert\,\mathcal{K}} \coloneqq \sum_{k=1}^{\mathcal{K}}E_{\pm 2k}g^{2k} \,.
\end{align}
The convergence behavior of $R_{0}\Omega_{\pm}$ and $\frac{L_{\pm}}{R_{0}}$ is computed analogously.
Generally, the quantities corresponding to prograde circular motion exhibit a significantly better convergence behavior than those of retrograde circular motion. It should be noted that $\Omega_{\pm}$ and $L_{\pm}$ are only exact up to ninth order.\footnote{Expanding a quantity that itself involves series expansions up to a certain order does not necessarily lead to a series expansion that is exact up to this order.}

\section{Conclusions and outlook}
\label{sec:conclusions}

Within general relativity various aspects of circular motion of uncharged test particles in the equatorial plane of the charged rotating disc of dust were analyzed.
Angular velocity, specific angular momentum and specific energy of the test particles were discussed in dependence of the specific charge and the relativity parameter. Also the radii of photon, marginally bound and marginally stable orbits were determined. In addition, formulae for angular velocity, specific angular momentum and specific energy of the test particles were derived that can be applied to the (exterior) spacetime of any isolated, axisymmetric, stationary, reflection symmetric, charged, rotating body.

It can be concluded that for sufficiently large radii (at least \mbox{$\rho>\rho_{+}^{\text{ms}}$}), the qualitative behavior of equatorial circular motion around the disc is very similar to that in the exterior spacetime of a Kerr-Newman black hole. In the close vicinity of the disc, however, circular motion shows a very distinct behavior from the one near the event horizon of a black hole.\footnote{This is consistent with the conclusion drawn in \cite{Ansorg1998} about equatorial circular motion around the uncharged rotating disc of dust.}

Both methods presented can be generalized to charged test particles. General formulae to be obtained for charged test particles can then also be applied to the charged rotating disc of dust. 
Similar to the transition from neutral to charged test particles in the case of the Kerr-Newman spacetime, this should reveal all kinds of new facets.

Furthermore, both methods can also be adapted to null geodesics. In this way also circular photon orbits in the equatorial plane of the disc can be studied in detail.
Another possibility is to investigate circular motion (of either neutral/charged test particles or photons) in the interior of the disc. Without further ado, general formulae for the exterior can also be used for the interior spacetime of the disc, since the metric of the disc is globally written in terms of Weyl-Lewis-Papapetrou coordinates.
Clearly, circular motion in the interior of the disc is quite different from that in the exterior.

\begin{acknowledgments}
	This work has been funded by the Deutsche
	Forschungsgemeinschaft (DFG) under Grant No.
	406116891 within the Research Training Group RTG
	2522/1. 
	The author would like to thank Reinhard Meinel and Lars Maiwald for valuable discussions
	and \mbox{Martin} Breithaupt for the provided disc solution in terms of the post-Newtonian expansion up to tenth order.
\end{acknowledgments}

\appendix

\section{Post-Newtonian expansion of the normalized angular velocity, the normalized specific angular momentum and the specific energy of  neutral test particles}\label{secB}
\sectionmark{Post-Newtonian expansion of $R_{0}\Omega_{\pm}$, $\frac{L_{\pm}}{R_{0}}$ and $E_{\pm}$}

The post-Newtonian expansion of the normalized angular velocity $R_{0}\Omega_{\pm} $ (up to first order), the normalized specific angular momentum $\frac{L_{\pm}}{R_{0}}$ (up to first order) and the specific energy $E_{\pm}$ (up to second order) of uncharged test particles moving along equatorial circular geodesics is as follows:
\onecolumngrid
\begin{align*}
	R_{0}\Omega_{+}
	&= \frac{\sqrt{\left(\pi - 2\arctan\! \left(\nu \right)\right) \left(1 + \nu^2\right) - 2\nu}\, g}{\sqrt{\pi}\, \sqrt{1+\nu^{2}}} \\
	&\quad+\frac{1}{\pi^{2} \sqrt{\left(\pi - 2\arctan\! \left(\nu \right)\right) \left(1 + \nu^2\right) - 2\nu}\, \sqrt{1+\nu^{2}}} \\
	&\quad\cdot\left[\left(\frac{13}{4} \left(1+\nu^{2}\right) \left(\left(\epsilon^{2}-\frac{36}{13}\right) \nu^{2}+\frac{23 \epsilon^{2}}{39}+\frac{2}{13}\right) \arctan\! \left(\nu \right)-9 \nu^{3}-\frac{11 \nu}{2} \right.\right. \\
	&\quad\left.\left.+\,\frac{13 \epsilon^{2} \nu^{3}}{4}+\frac{15 \nu  \,\epsilon^{2}}{4}\right) \pi^{\frac{3}{2}}-\frac{15}{8} \left(1+\nu^{2}\right) \left(\left(\epsilon^{2}-\frac{8}{5}\right) \nu^{2}+\frac{17 \epsilon^{2}}{45}-\frac{2}{15}\right) \pi^{\frac{5}{2}} \right. \\
	&\quad\left.+\,\left(\arctan \! \left(\nu \right) \left(1+\nu^{2}\right)+\nu\right) \sqrt{\pi}\, \left(\left(\nu^{2} \left(\epsilon^{2}+6\right)-\epsilon^{2}-2\right) \arctan \! \left(\nu \right)    + \left(\epsilon^{2}+6\right) \nu \right)\vphantom{\frac{2}{15}} \right. \\
	&\quad\left.-\,3 \sqrt{1-\epsilon^{2}}\, \sqrt{\left(\pi - 2\arctan\! \left(\nu \right)\right) \left(1 + \nu^2\right) - 2\nu}\, \sqrt{1+\nu^{2}} \left(\vphantom{\frac{2}{15}}\pi  \,\nu^{2} -2 \arctan \! \left(\nu \right) \nu^{2} \right.\right. \\
	&\quad\left.\left.+\,\frac{\pi}{3}-2 \nu -\frac{2 \arctan \left(\nu \right)}{3}\right) \pi 
	+ \frac{5\pi^{3/2}}{6}\left(\left(\pi - 2\arctan\! \left(\nu \right)\right) \left(1 + \nu^2\right) - 2\nu\right)\right] g^{3}
	+ \mathcal{O}\left(g^{5}\right) \,,
\end{align*}
\begin{align*}
	R_{0}\Omega_{-}
	&=-\frac{\sqrt{\left(\pi - 2\arctan\! \left(\nu \right)\right) \left(1 + \nu^2\right) - 2\nu}\, g}{\sqrt{\pi}\, \sqrt{1+\nu^{2}}} \\
	&\quad+\frac{1}{\pi^{2} \sqrt{\left(\pi - 2\arctan\! \left(\nu \right)\right) \left(1 + \nu^2\right) - 2\nu}\, \sqrt{1+\nu^{2}}} \\
	&\quad\cdot\left[- \left(\left(\frac{13}{4} \left(1+\nu^{2}\right) \left(\left(\epsilon^{2}-\frac{36}{13}\right) \nu^{2}+\frac{23 \epsilon^{2}}{39}+\frac{2}{13}\right) \arctan\!\left(\nu \right)-9 \nu^{3}-\frac{11 \nu}{2} \right.\right.\right. \\
	&\quad\left.\left.\left.+\,\frac{13 \epsilon^{2} \nu^{3}}{4}+\frac{15 \nu  \,\epsilon^{2}}{4}\right) \pi^{\frac{3}{2}}-\frac{15}{8} \left(1+\nu^{2}\right) \left(\left(\epsilon^{2}-\frac{8}{5}\right) \nu^{2}+\frac{17 \epsilon^{2}}{45}-\frac{2}{15}\right) \pi^{\frac{5}{2}} \right.\right. \\
	&\quad\left.\left.+\left(\arctan \! \left(\nu \right) \left(1+\nu^{2}\right) +\nu\right) \sqrt{\pi}\, \left(\left(\nu^{2} \left(\epsilon^{2}+6\right)-\epsilon^{2}-2\right) \arctan \! \left(\nu \right)   + \left(\epsilon^{2}+6\right) \nu \right)\vphantom{\frac{2}{15}}\right)  \right. \\
	&\quad\left.-\,3 \sqrt{1-\epsilon^{2}}\, \sqrt{\left(\pi - 2\arctan\! \left(\nu \right)\right) \left(1 + \nu^2\right) - 2\nu}\, \sqrt{1+\nu^{2}} \left(\vphantom{\frac{2}{15}}\pi  \,\nu^{2} -\,2 \arctan \! \left(\nu \right) \nu^{2} \right.\right. \\
	&\quad\left.\left.+\frac{\pi}{3}-2 \nu -\frac{2 \arctan \left(\nu \right)}{3}\right) \pi 
	- \frac{5\pi^{3/2}}{6}\left(\left(\pi - 2\arctan\! \left(\nu \right)\right) \left(1 + \nu^2\right) - 2\nu\right)\right] g^{3}
	+ \mathcal{O}\left(g^{5}\right) \,,
\end{align*}
\begin{align*}
	\frac{L_{+}}{R_{0}}
	&= \frac{\sqrt{\left(\pi - 2\arctan\! \left(\nu \right)\right) \left(1 + \nu^2\right) - 2\nu}\, \sqrt{1+\nu^{2}}\, g}{\sqrt{\pi}} \\
	&\quad-\frac{3}{2\pi^{\frac{27}{2}} \left(\left(\pi - 2\arctan\! \left(\nu \right)\right) \left(1 + \nu^2\right) - 2\nu \right)} \\
	&\quad\cdot\left[\frac{5 \pi^{12}}{4} \left(-\frac{8}{15} \left(\nu^{2} \left(\epsilon^{2}+2\right)-\epsilon^{2}+6\right) \left(1+\nu^{2}\right) \arctan^{2}\! \left(\nu \right)+\left(\pi  \left(-\frac{26 \epsilon^{2}}{15}+\frac{8}{3}\right) \nu^{4} \right.\right.\right. \\
	&\quad\left.\left.\left.+\,\left(-\frac{32}{15}-\frac{16 \epsilon^{2}}{15}\right) \nu^{3}+\pi  \left(\frac{20}{3}-\frac{124 \epsilon^{2}}{45}\right) \nu^{2}-\frac{64 \nu}{15}+\pi  \left(4-\frac{46 \epsilon^{2}}{45}\right)\right) \arctan \left(\nu \right) \right.\right. \\
	&\quad\left.\left.+\,\pi^{2} \left(\epsilon^{2}-\frac{16}{15}\right) \nu^{4}+\pi  \left(-\frac{26 \epsilon^{2}}{15}+\frac{8}{3}\right) \nu^{3}+\left(\left(-\frac{34}{15}+\frac{62 \epsilon^{2}}{45}\right) \pi^{2}-\frac{16}{15}-\frac{8 \epsilon^{2}}{15}\right) \nu^{2} \right.\right. \\
	&\quad\left.\left.+\,\left(-2 \epsilon^{2}+4\right) \pi  \nu +\pi^{2} \left(\frac{17 \epsilon^{2}}{45}-\frac{6}{5}\right)\right) \sqrt{1+\nu^{2}}\, \sqrt{\left(\pi - 2\arctan\! \left(\nu \right)\right) \left(1 + \nu^2\right) - 2\nu} \right. \\
	&\quad\left.+\,\left(4 \left(\left(1+\nu^{2}\right)^{2} \arctan \! \left(\nu \right)+\nu^{3}+\frac{5 \nu}{3}\right) \left(\left(1+\nu^{2}\right) \arctan \! \left(\nu \right)+\nu \right) \pi^{\frac{25}{2}} \right.\right. \\
	&\quad\left.\left.+\,\left(-4 \left(1+\nu^{2}\right)^{2} \arctan \! \left(\nu \right)+\pi  \,\nu^{4}+2 \pi  \,\nu^{2}-4 \nu^{3}+\pi -\frac{16 \nu}{3}\right) \pi^{\frac{27}{2}} \left(1+\nu^{2}\right)\right) \sqrt{1-\epsilon^{2}} \right. \\
	&\quad\left.+\, \frac{5\pi^{13}}{9}\left(\left(\pi - 2\arctan\! \left(\nu \right)\right) \left(1 + \nu^2\right) - 2\nu\right)^{3/2}\sqrt{1+\nu^2}   \right] g^{3}
	+ \mathcal{O}\left(g^{5}\right) \,,
\end{align*}
\begin{align*}
	\frac{L_{-}}{R_{0}} 
	&=-\frac{\sqrt{\left(\pi - 2\arctan\! \left(\nu \right)\right) \left(1 + \nu^2\right) - 2\nu}\, \sqrt{1+\nu^{2}}\, g}{\sqrt{\pi}} \\
	&\quad-\frac{3}{2\pi^{\frac{27}{2}} \left(\left(\pi - 2\arctan\! \left(\nu \right)\right) \left(1 + \nu^2\right) - 2\nu \right)} \\
	&\quad\cdot\left[-\frac{5 \pi^{12}}{4} \left(-\frac{8}{15} \left(\nu^{2} \left(\epsilon^{2}+2\right)-\epsilon^{2}+6\right) \left(1+\nu^{2}\right) \arctan^{2}\! \left(\nu \right)+\left(\pi  \left(-\frac{26 \epsilon^{2}}{15}+\frac{8}{3}\right) \nu^{4} \right.\right.\right. \\
	&\quad\left.\left.\left.+\,\left(-\frac{32}{15}-\frac{16 \epsilon^{2}}{15}\right) \nu^{3}+\pi  \left(\frac{20}{3}-\frac{124 \epsilon^{2}}{45}\right) \nu^{2}-\frac{64 \nu}{15}+\pi  \left(4-\frac{46 \epsilon^{2}}{45}\right)\right) \arctan \left(\nu \right) \right.\right. \\
	&\quad\left.\left.+\,\pi^{2} \left(\epsilon^{2}-\frac{16}{15}\right) \nu^{4}+\pi  \left(-\frac{26 \epsilon^{2}}{15}+\frac{8}{3}\right) \nu^{3}+\left(\left(-\frac{34}{15}+\frac{62 \epsilon^{2}}{45}\right) \pi^{2}-\frac{16}{15}-\frac{8 \epsilon^{2}}{15}\right) \nu^{2} \right.\right. \\
	&\quad\left.\left.+\,\left(-2 \epsilon^{2}+4\right) \pi  \nu +\pi^{2} \left(\frac{17 \epsilon^{2}}{45}-\frac{6}{5}\right)\right) \sqrt{1+\nu^{2}}\, \sqrt{\left(\pi - 2\arctan\! \left(\nu \right)\right) \left(1 + \nu^2\right) - 2\nu} \right. \\
	&\quad\left.+\,\left(4 \left(\left(1+\nu^{2}\right)^{2} \arctan \! \left(\nu \right)+\nu^{3}+\frac{5 \nu}{3}\right) \left(\left(1+\nu^{2}\right) \arctan \! \left(\nu \right)+\nu \right) \pi^{\frac{25}{2}} \right.\right. \\
	&\quad\left.\left.+\,\left(-4 \left(1+\nu^{2}\right)^{2} \arctan \! \left(\nu \right)+\pi  \,\nu^{4}+2 \pi  \,\nu^{2}-4 \nu^{3}+\pi -\frac{16 \nu}{3}\right) \pi^{\frac{27}{2}} \left(1+\nu^{2}\right)\right) \sqrt{1-\epsilon^{2}}\right. \\
	&\quad\left.-\, \frac{5\pi^{13}}{9}\left(\left(\pi - 2\arctan\! \left(\nu \right)\right) \left(1 + \nu^2\right) - 2\nu\right)^{3/2}\sqrt{1+\nu^2}   \right] g^{3}
	+ \mathcal{O}\left(g^{5}\right) \,,
\end{align*}
\begin{align*}
	E_{+} 
	&= 1+\frac{\nu  \left(\pi  \nu -2 \arctan \! \left(\nu \right) \nu -2\right) g^{2}}{\pi} \\
	&\quad-\frac{3}{\sqrt{1+\nu^{2}}\, \pi^{\frac{11}{2}}} \left[\frac{\pi^{4}}{3} \sqrt{\left(\pi - 2\arctan\! \left(\nu \right)\right) \left(1 + \nu^2\right) - 2\nu}\, \left(1+\nu^{2}\right) \right. \\
	&\quad\left.\cdot\,\left(\left(-6 \nu^{2}-2\right) \arctan\! \left(\nu \right)+3 \pi  \,\nu^{2}+\pi -6 \nu \right) \sqrt{1-\epsilon^{2}}+\frac{15}{16} \left(\left(\left(-\frac{16}{9}-\frac{8 \epsilon^{2}}{15}\right) \nu^{4} \right.\right.\right. \\
	&\quad\left.\left.\left.+\,\left(-\frac{32}{15}+\frac{16 \epsilon^{2}}{45}\right) \nu^{2}-\frac{16}{15}+\frac{8 \epsilon^{2}}{45}\right) \arctan^2\!\left(\nu \right)+\left(\left(\left(\frac{152}{45}-\frac{26 \epsilon^{2}}{15}\right) \nu^{4} \right.\right.\right.\right. \\
	&\quad\left.\left.\left.\left.+\,\left(\frac{64}{15}-\frac{68 \epsilon^{2}}{27}\right) \nu^{2}-\frac{58 \epsilon^{2}}{135}+\frac{56}{45}\right) \pi -\frac{16 \nu }{15} \left(\left(\frac{10}{3}+\epsilon^{2}\right) \nu^{2}-\frac{\epsilon^{2}}{3}+2\right)\right) \arctan\! \left(\nu \right) \right.\right. \\
	&\quad\left.\left.+\,\left(1+\nu^{2}\right) \left(\left(-\frac{56}{45}+\epsilon^{2}\right) \nu^{2}+\frac{23 \epsilon^{2}}{135}-\frac{16}{45}\right) \pi^{2}+\left(-\frac{26}{15} \epsilon^{2} \nu^{3}-\frac{214}{135} \nu  \,\epsilon^{2}+\frac{152}{45} \nu^{3}+\frac{8}{3} \nu \right) \pi \right.\right. \\
	&\quad\left.\left.-\,\frac{8}{15} \left(\frac{10}{3}+\epsilon^{2}\right) \nu^{2}\right) \pi^{\frac{7}{2}} \sqrt{1+\nu^{2}}\right] g^{4} + \mathcal{O}\left(g^{6}\right) \,,
\end{align*}
\begin{align*}
	E_{-}
	&= 1+\frac{\nu  \left(\pi  \nu -2 \arctan \! \left(\nu \right) \nu -2\right) g^{2}}{\pi} \\
	&\quad-\frac{45 }{16 \sqrt{1+\nu^{2}}\, \pi^{\frac{11}{2}}}\left[-\frac{16}{15} \sqrt{1-\epsilon^{2}}\, \left(1+\nu^{2}\right) \left(\pi  \,\nu^{2}-2 \arctan\! \left(\nu \right) \nu^{2}+\frac{\pi}{3}-2 \nu \right.\right. \\
	&\quad\left.\left.-\,\frac{2 \arctan\! \left(\nu \right)}{3}\right) \pi^{4} \sqrt{\left(\pi - 2\arctan\! \left(\nu \right)\right) \left(1 + \nu^2\right) - 2\nu} \right. \\
	&\quad\left.+\,\left(\left(\left(-\frac{16}{9}-\frac{8 \epsilon^{2}}{15}\right) \nu^{4}+\left(-\frac{32}{15}+\frac{16 \epsilon^{2}}{45}\right) \nu^{2}-\frac{16}{15}+\frac{8 \epsilon^{2}}{45}\right) \arctan^{2} \! \left(\nu \right) \right.\right. \\
	&\quad\left.\left.+\,\left(\left(\left(\frac{152}{45}-\frac{26 \epsilon^{2}}{15}\right) \nu^{4}+\left(\frac{64}{15}-\frac{68 \epsilon^{2}}{27}\right) \nu^{2}-\frac{58 \epsilon^{2}}{135}+\frac{56}{45}\right) \pi \right.\right.\right. \\
	&\quad\left.\left.\left.-\,\frac{16 \nu}{15}  \left(\left(\frac{10}{3}+\epsilon^{2}\right) \nu^{2}-\frac{\epsilon^{2}}{3}+2\right)\right) \arctan \! \left(\nu \right)+\left(1+\nu^{2}\right) \left(\left(-\frac{56}{45}+\epsilon^{2}\right) \nu^{2}+\frac{23 \epsilon^{2}}{135}-\frac{16}{45}\right) \pi^{2}\right.\right. \\
	&\quad\left.\left.+\,\left(-\frac{26}{15} \epsilon^{2} \nu^{3}-\frac{214}{135} \nu  \,\epsilon^{2}+\frac{152}{45} \nu^{3}+\frac{8}{3} \nu \right) \pi -\frac{8}{15} \left(\frac{10}{3}+\epsilon^{2}\right) \nu^{2}\right) \pi^{\frac{7}{2}} \sqrt{1+\nu^{2}}\right] g^{4} + \mathcal{O}\left(g^{6}\right) \,.
\end{align*}
\newline

\twocolumngrid



\bibliography{cm-bibliography}

\end{document}